\documentclass[final,12p,times,twocolumn]{elsarticle}

 \usepackage{graphicx}
\usepackage{amssymb}
\usepackage{amsthm}

\usepackage[nodots]{numcompress}

\biboptions{sort&compress}
\journal{Carbon}
\begin{document}

\begin{frontmatter}



\title{Raman spectroscopic characterization of stacking configuration and interlayer coupling of twisted multilayer graphene grown by chemical vapor deposition}


\author[1]{Jiang-Bin Wu}
\author[2]{Huan Wang}
\author[1]{Xiao-Li Li}
\author[2]{Hailin Peng}
\author[1]{Ping-Heng Tan\corref{cor1}}
\ead{phtan@semi.ac.cn}
\cortext[cor1]{Corresponding author.}

\address[1]{State Key Laboratory of Superlattices and
Microstructures, Institute of Semiconductors, Chinese Academy of
Sciences, Beijing 100083, China}
\address[2]{Center for Nanochemistry, Beijing National Laboratory for Molecular Sciences, Key Laboratory for the Physics and Chemistry of Nanodevices, College of Chemistry and Molecular Engineering, Peking University, Beijing 100871, China}

\begin{abstract}
Multilayer graphene (MLG) grown by chemical vapor deposition (CVD) is a promising material for electronic and optoelectronic devices. Understanding the stacking configuration and interlayer coupling of MLGs is technologically relevant and of importance for the device applications. Here, we reported a kind of twisted MLGs (tMLGs), which are formed by stacking one graphene monolayer on the top of AB-stacked MLG by rotating a certain angle between them. The twist angle of tMLGs are identified by the twist-related modes, R and R'. With increasing the total layer number of N, the observed interlayer shear modes in the tMLG flake always follow those of AB-stacked (N-1)LG, while the observed interlayer breathing modes always follow those of AB-stacked NLG, independent of its twist angle. The layer breathing coupling of the tMLGs is almost identical to that of mechanically-exfoliated tMLGs, which demonstrates the high quality of MLGs grown by CVD. This study provides an applicable approach to probe the stacking configuration and interlayer coupling of MLGs grown by CVD or related methods. This work also demonstrates the possibility to grow MLG flakes with a fixed stacking configuration, $e.g.$, t(1+$n$)LG, by the CVD method.
\end{abstract}

%\begin{keyword}


%\end{keyword}

\end{frontmatter}

% \linenumbers

\newpage

\section{Introduction}
\label{}



%% References


Large-area multilayer graphene (MLG) grown by chemical vapor deposition (CVD), with transparency and high electrical conductivity and flexibility, is considered as a candidate for transparent and conducting electrodes, which can be used in touch screen panels, organic light-emitting diodes and solar cells\cite{kim-Nature-2009-large,li-2009-large,gomez-2010-continuous,bae-2010-roll,xue-2011-scanning,wang-2011-interface,wang-2011-new,han-2012-extremely}. The plane-to-plane (vertical) conductivity, determined by the interlayer coupling of MLG, is the bottleneck of improving the overall conductivity, due to the series resistance effect\cite{liu-2014-giant,lee-2014-large}.  Therefore, it's important to produce and characterize the MLG with an interlayer coupling close to that of graphite. The CVD method tends to produce twisted bilayer graphene (2LG)\cite{Lee-NL-2010-wafer,yan2011growth,nie2011growth,Yan-NL-2011},

Twisted 2LG (t2LG) is non-AB stacked bilayer graphene in which one graphene monolayer sheet rotates by a certain angle ($\theta_t$) relative to the other\cite{lui-2010-NL-imaging,dos-PRL-2007-graphene}.  t2LG brings about a series of novel physical properties, $e.g.$, lower Fermi velocity than single layer graphene (SLG)\cite{dos-PRL-2007-graphene,trambly-NL-2010-localization} and $\theta_t$-dependent optical absorption\cite{havener-NL-2012-angle,Moon_PRB_2013}, and can be employed as photoelectric detector\cite{yeh-2014-gating} and pressure sensors\cite{nguyen2014strain,nguyen-2014s-train}. Similarly, by assembling m-layer (mLG, m$>$1) and n-layer (nLG, n$\ge$1) flakes, a (m+n)-system can be formed, which is denoted as t(m+n)LG, a kind of twisted MLG (tMLG). In general, for a given total layer number $N$ (with $N$=$n$+$m$+...), t($n$+$m$+...)LG is assumed to denote the $N$ layer graphene, which is stacked by each AB-stacked $n$,$m$...LG with twist angles between them. The stacking configuration information in a t(n+m+...)LG includes the layer number  of each constituent and the stacking way between two adjacent constituents.\cite{wu-nc-2014-resonant} tMLG (layer number $>$2) also exhibits a series of novel physical properties\cite{wu-nc-2014-resonant}, similar to the case of t2LG. One can easily grow t2LG or AB-stacked 2LG (AB-2LG) by CVD to control its flake size and stacking configuration. However, it will become more and more important to grow graphene flakes by CVD to create tMLG on demand with properties determined by the interlayer interaction and stacking configuration. When more graphene layers are grown by CVD, it's still a challenge to characterize the stacking configuration and interlayer coupling of CVD-grown MLG (CVD-MLG) directly and nondestructively.

Raman spectroscopy is one of the most useful characterization techniques in graphene\cite{FerrariNN,ferrari-SSC-2007-raman}.The phonon vibration modes of AB-stacked MLG (AB-MLG) can be divided into the in-plane vibration modes (like G and 2D modes) and out-plane vibration modes (like the shear (C) mode). Besides G and 2D modes, two additional R and R$'$ modes, which are from the TO and LO phonon branches respectively, selected by a twist vector, are detected in t(n+m)LG\cite{Carozo-2011-NL-Raman,campos-2013-raman,Ado2013SSC,wu-nc-2014-resonant}, and can be a signature to distinguish the stacking configuration between twist and AB stacking. The layer number of AB-MLG can be determined by the 2D\cite{ferrari-PRL-2006-raman} and C\cite{tan-2012-NM-shear} modes. The 2D peak profile of the AB-MLG is layer-number dependent\cite{ferrari-PRL-2006-raman}, because the electronic structures of the AB-MLG are distinct\cite{latil-2006-charge} and the 2D peak profile is related to its electronic structures by the double resonant process\cite{thomsen}. However, the twist angle dependent electronic structures\cite{trambly-NL-2010-localization} in the tMLGs lead to much more complicated profile of 2D peaks\cite{kim-prl-2012-raman,Cong-prb-2014}, which aren't suitable to identify the layer number of tMLGs. Moreover, twisting would block the interlayer shear coupling, resulting in the localization of the C modes in the AB-stacked constituent\cite{wu-nc-2014-resonant}. Fortunately, the interlayer breathing coupling remains almost constant at the twisted interface, and the layer breathing (LB) modes (LBMs), which can not be measured in MLG, can be detected in tMLG under the resonant condition.\cite{Wu_LBM_ACSnano} Therefore, by monitoring the C and LB modes, the stacking configuration between adjacent constituents in tMLG can be determined. Moreover, the C and LB modes correspond to interlayer vibrations, whose frequencies are directly related to the strength of interlayer coupling strengh\cite{tan-2012-NM-shear,zhang-PhysRevB-2013}. Therefore, it is necessary to extend the method of identifying stacking configuration and interlayer coupling strength in mechanically-exfoliated tMLGs (ME-tMLGs) by ultralow-frequency (ULF) Raman spectroscopy to MLG flakes grown by CVD.

Here, we employ CVD method to prepare MLG flakes containing different layer numbers ranging from 1 to 7, whose layer number is determined by Rayleigh imaging\cite{Casiraghi-nano-2007}. The CVD-MLGs are identified as twisted (1+$n$)LG by the R and R$'$ modes. In the ULF region, with increasing the total layer number of N, the observed C modes in the tMLG with a definite $\theta_t$ always follow those of AB-(N-1)LG, while its observed LBMs always follow those of AB-NLG. The tMLG with a definite total layer number of N exhibits similar spectral features ($e.g.$, the number and frequency of the C and LB modes), independent of $\theta_t$. Based on the frequencies of C and LB modes, we revealed that the interlayer C and LB couplings in tMLGs prepared by both CVD method and self-folding during the mechanical exfoliation process are almost identical to each other, respectively, which indicates the high crystal quality of tMLGs grown by the CVD method.

\section{Experimental}

\subsection{Sample preparation.}MLG flacks were grown on annealed copper substrate (Alfa Aesar) in a homemade low-pressure chemical vapor deposition (LPCVD) system. Before growth, the copper foil was under pre-annealed treatment at 1020 $\ensuremath{^{\circ}}$C under 100 sccm H$_2$ with a pressure of about 200 Pa for 30min to reduce oxide at the surface of copper. H$_2$ flow was changed into 600 sccm and 1 sccm methane was introduced while keeping the temperature constant of 1020 $\ensuremath{^{\circ}}$C. After 40 min growth, the furnace was cooled down to room temperature. MLG flacks grown on copper were transferred onto Si substrate with (90 nm) SiO$_2$ thickness with the aid of Poly (methylmethacrylate).

\subsection{Raman measurements.} Raman spectra are measured in back-scattering at room temperature with a Jobin-Yvon HR800 Raman system, equipped with a liquid-nitrogen-cooled charge-coupled device (CCD), a 100$\times$ objective lens (NA=0.90) and several gratings. The excitation wavelength is 785 nm from a Ti:Saphire laser, 633 nm from a He-Ne laser, 530 and 568 nm from a Kr$^+$ laser, and 488, 466 nm from an Ar$^+$ laser. The resolution of the Raman system at 633 nm is 0.35 cm$^{-1}$ per CCD pixel. Plasma lines are removed from the laser signals, using a BragGrate Bandpass filters. Measurements down to 5 cm$^{-1}$ for each excitation are enabled by three BragGrate notch filters with optical density 3 and with FWHM=5-10 cm$^{-1}.$\cite{tan-2012-NM-shear} Both BragGrate bandpass and notch filters are produced by OptiGrate Corp. The typical laser power of $\sim$0.5mW is used to avoid sample heating.

\section{Results and Discussions}

\begin{figure*}[!htb]
\centerline{\includegraphics[width=140mm,clip]{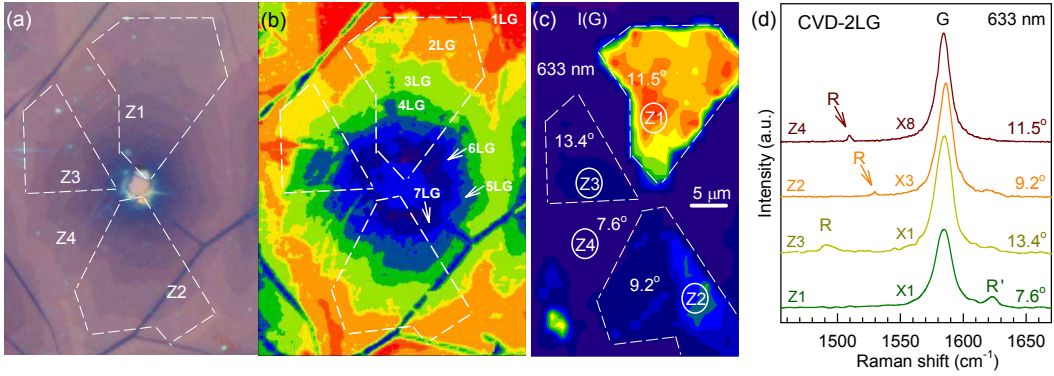}}
\caption{\textbf{Characterization of CVD-grown MLG.} ({\bf a}) Optical image of a CVD-MLG flake. ({\bf b}) Contrast mapping of Rayleigh scattering of the CVD-MLG flake shown in (a) under the 532 nm excitation, from which the layer numbers of the MLG flakes in different regions are determined and marked. ({\bf c}) Raman image of the G mode intensity under the 633 nm excitation for the corresponding MLG flake. The zones with different G mode intensities are indicated by white dash lines and marked as Z1, Z2, Z3 and Z4, respectively. The images in (a) (b) (c) are with the same scale. ({\bf d}) Raman spectra of CVD-2LG flakes in different zones in (c). The scale factor of each spectrum is marked.} \label{Fig1}
\end{figure*}

The optical image of a typical CVD-grown polycrystalline MLG flake is shown in Fig. 1(a). The layer number of the MLG flake is identified by Rayleigh scattering\cite{Casiraghi-nano-2007}. Fig. 1(b) shows contrast mapping of Rayleigh scattering of a MLG flake obtained by 532 nm excitation. Rayleigh scattering signals are processed by the mean intensity of the substrate: $I(contrast)$=($I(sub.)$-$I(nLG)$)/$I(sub.)$, where $I(sub.)$ is the Rayleigh intensity from bara substrate and $I(nLG)$ is the Rayleigh intensity from the MLG flakes laid on the substrate. The region of each graphene flake with a definite layer number can be distinguished by I(contrast) or the color in the contrast mapping of Rayleigh scattering, as marked in the image. The CVD-2LG tends to be t(1+1)LG.\cite{Lee-NL-2010-wafer,yan2011growth,nie2011growth,Yan-NL-2011} The Van Hove singularities (VHS) in the electronic joint density of states (JDOSs) of all the optically allowed transitions\cite{wu-nc-2014-resonant} in t(1+1)LG would enhance the intensity of Raman signals, when the excitation energy matches the VHS energy, which is related to $\theta_t$. The relationship  between VHS and $\theta_t$ can be estimated by this formula\cite{kim-prl-2012-raman} of ${E}_{VHS}\approx4\pi{\theta}_{t}\hbar{v}_{f}/3a$, where $a$ is the lattice constant of graphene (2.46 \AA), $\hbar$ is the reduced Planck's constant, and ${v}_{f}$ is the Fermi velocity of SLG (${10}^{6} m/s$). The mapping of the G mode intensity (I(G)) of the CVD-MLG flake under the excitation of 633 nm (1.96 eV) is shown in Fig. 1(c). The CVD-MLG flake can be distinguished with different zones by different enhancement levels of I(G). Each zone is indicated by the white dash lines, and marked as Z1, Z2, Z3 and Z4, respectively. We also marked each zone in Figs. 1(a) and 1(b). The Raman spectrum of the CVD-2LG flake from each zone is shown in Fig. 1(d). The so-called R and R$'$ modes\cite{Carozo-2011-NL-Raman,wu-nc-2014-resonant} are observed in all the spectra, which suggests that the CVD-2LG flakes are t(1+1)LG. Those modes can be used to distinguish twist stacking in tMLG with a $\theta_t$ from AB stacking. $\theta_t$ of graphene flakes in the each zone can be indicated by the position of R and R$'$ modes,\cite{Carozo-2011-NL-Raman,Carozo-peb-2013,Ado2013SSC,wu-nc-2014-resonant} and are labeled in the mapping image in Fig. 1(c). The t(1+1)LG in the Z1 zone with a $\theta_t$ of 11.5$^{\circ}$ corresponds to an optimal excitation energy of 2.04 eV, which is close to the excitation energy of 1.96 eV, leading to a strong enhancement of I(G) in the Z1 zone. $\theta_t$ is more close to 10.8$^{\circ}$, the I(G) enhancement would be more significant under the 1.96 eV (633 nm) excitation.\cite{wu-nc-2014-resonant} We notice that the G mode intensity of the CVD-MLG flake, whose layer number is $>$2, is also enhanced, which indicates that the corresponding CVD-MLG flakes are tMLGs.

\begin{figure*}[!htb]
\centerline{\includegraphics[width=110mm,clip]{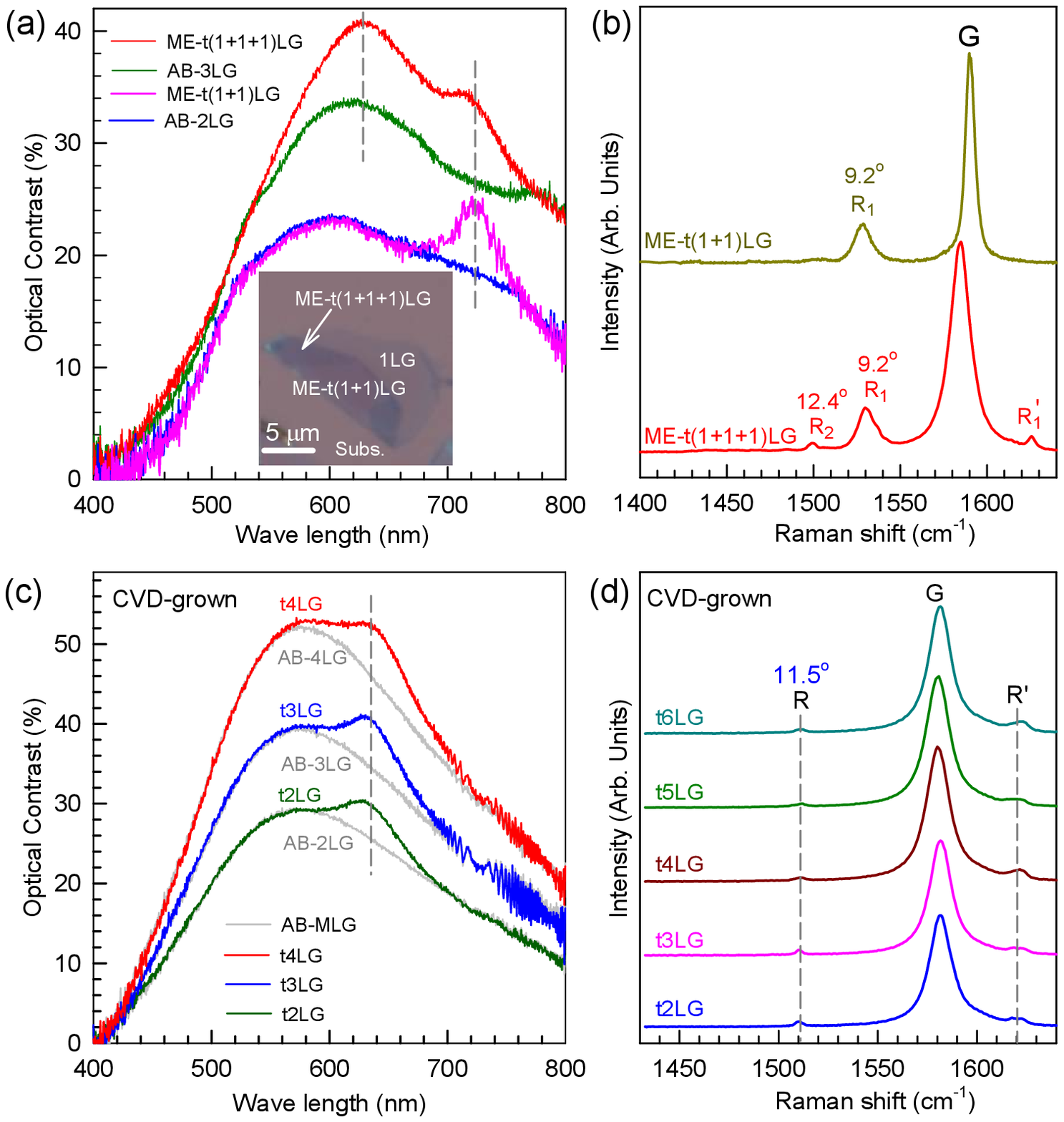}}
\caption{\textbf{Characterization of CVD-grown t$N$LG.} ({\bf a}) Optical contrast of ME-t(1+1)LG and ME-t(1+1+1)LG. The optical contrast of AB-2LG, ME-t(1+1)LG, AB-3LG and ME-t(1+1+1)LG are plotted by blue, pink, green and red lines, respectively. One adsorption peak is observed in ME-t(1+1)LG, and two adsorption peaks raise in ME-t(1+1+1)LG. The inset shows the optical image of the ME-t(1+1)LG and ME-t(1+1+1)LG flakes. ({\bf b}) Raman spectra of ME-t(1+1)LG and ME-t(1+1+1)LG. An R mode is observed in ME-t(1+1)LG. Two couples of R and R$'$ modes are observed in ME-t(1+1+1)LG. (c) Optical contrast of CVD-MLG flakes and that of the corresponding AB-MLGs (gray lines). In comparison with AB-MLGs, just one adsorption peak appears for all the CVD-MLGs at the same position of 640 nm. ({\bf d}) Raman spectra of CVD-grown t$N$LG ($N$=2,3,4,5,6). The same couple of R and R$'$ mode are observed in all the CVD-grown t$N$LG.} \label{Fig2}
\end{figure*}

To figure out the stacking way in the CVD-grown tMLGs, we prepared t(1+1)LG and t(1+1+1)LG by self-folding technique during the mechanical exfoliation process\cite{wu-nc-2014-resonant}, which are denoted as ME-t(1+1)LG and ME-t(1+1+1)LG in this paper, respectively. Based on their optical images shown in the inset to Fig. 2(a), the twisted interface between the bottom two layers of the ME-t(1+1+1)LG is identical to that of the ME-t(1+1)LG. The optical contrast of ME-t(1+1+1)LG and ME-t(1+1+1)LG is plotted in Fig.2(a) along with the corresponding AB-2LG and AB-stacked 3LG (AB-3LG). Compared with AB-2LG, twist in the ME-t(1+1)LG would raise an adsorption peak at $\sim$720 nm in the optical contrast, due to the presence of VHS in the JDOS.\cite{havener-NL-2012-angle,Moon_PRB_2013}, as shown in Fig. 2(a). Compared with AB-3LG, the adsorption peak at ~720 nm still exists in the optical contrast of ME-t(1+1+1)LG along with one additional adsorption peak at $\sim$620 nm. The same adsorption peak at $\sim$720 nm are found in ME-t(1+1)LG and ME-t(1+1+1)LG, resulting from the existence of the same twisted interface, as shown by their optical images in Fig. 2(a). Therefore, the additional adsorption peak at ~620 nm in ME-t(1+1+1)LG can be attributed to its formation of the second twisted interface after adding one graphene layer upon the ME-t(1+1)LG. This is further confirmed by the corresponding Raman spectra, as shown in Fig. 2(b). In the ME-t(1+1)LG, the R peak is observed at 1531 cm$^{-1}$, which corresponds to the twisted interface with a $\theta_t$ of 9.2$^{\circ}$. Two additional peaks are measured at 1500 cm$^{-1}$ and 1626 cm$^{-1}$ in the ME-t(1+1+1)LG. According to the phonon dispersion of SLG, the peak at 1626 cm$^{-1}$ is the R$'$ mode corresponding to the twisted interface between the bottom two graphene layers, which shares the same twisted vector with the R peak at 1531 cm$^{-1}$ of the ME-t(1+1)LG. Therefore, the peaks in 1531 and 1618 cm$^{-1}$ are denoted as R$_{1}$ and R$_{1}$$'$ respectively. The peak at 1500 cm$^{-1}$ indicating a 12.4$^{\circ}$ twist angle is denoted as R$_{2}$, corresponding to the other twisted interface between the top two layers of ME-t(1+1+1)LG. The above discussion demonstrates that the number of twisted interfaces in a tMLG flake can be determined from the adsorption peaks in the optical contrast difference between the tMLG flake and the corresponding AB-MLGs, and also from the sets of the R and R$'$ modes in the corresponding Raman spectra.

\begin{figure*}[!htb]
\centerline{\includegraphics[width=120mm,clip]{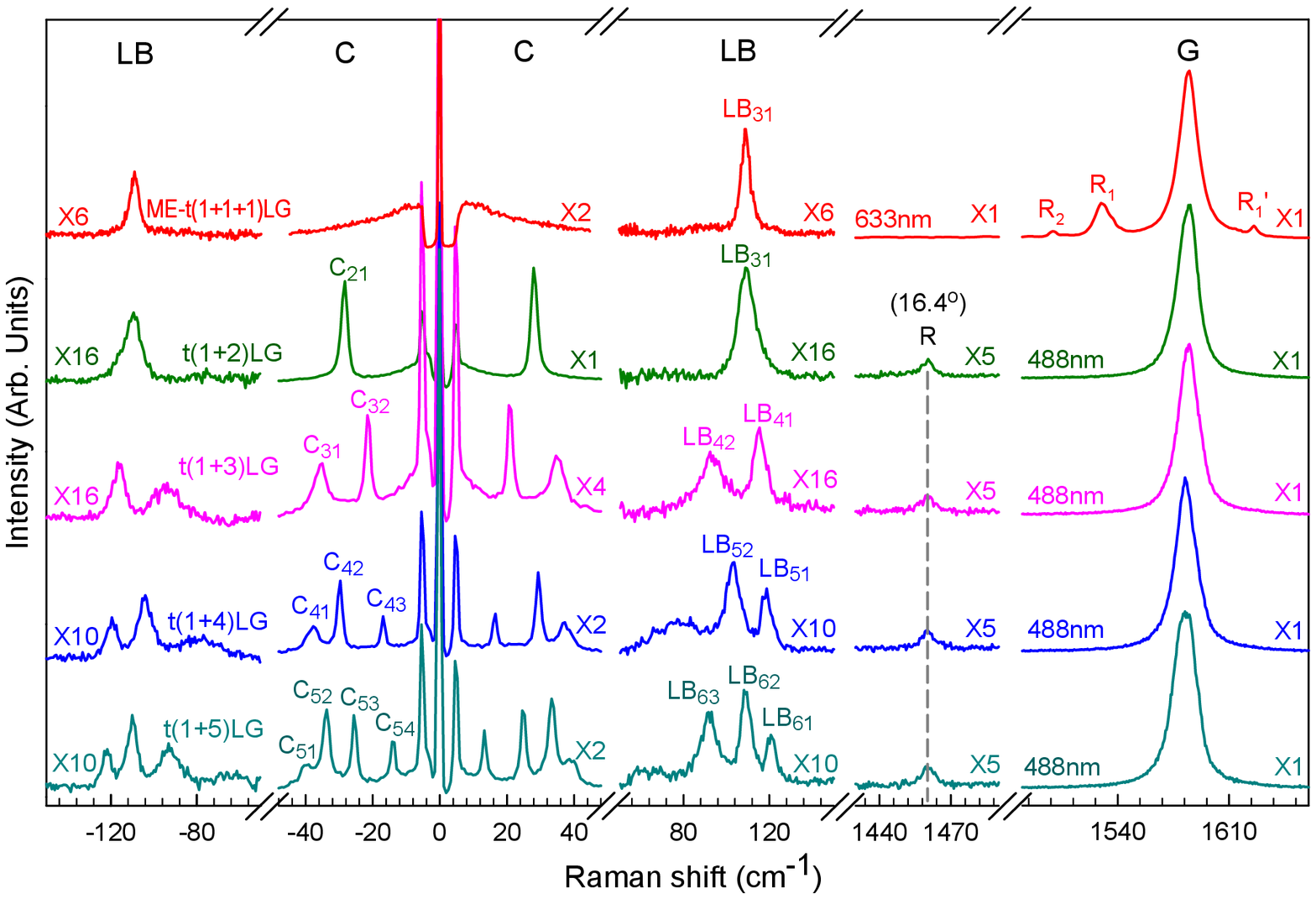}}
\caption{\textbf{Raman spectroscopy of CVD-grown t(1+$n$)LG.} Stokes/anti-Stokes Raman spectra in the C and LB mode region and Stokes spectra in the G spectral region for ME-t(1+1+1)LG and CVD-grown t(1+$n$)LG. The top one is the Raman spectra of t(1+1+1)LG excited by 633 nm. The rest are the Raman spectra of t(1+$n$)LG excited by 488 nm. The spectra are scaled and offset for clarity. The scaling factors of the individual spectra are shown. There are not any C modes observed in t(1+1+1)LG. For the CVD-grown sample, the layer number for each flake is 1+$n$. The C modes of each $n$LG are observed, while the LBMs of (1+$n$)LG are observed. Vertical dashed lines are a guide to the eye.} \label{Fig3}
\end{figure*}

Next, we checked a CVD-MLG flake containing t$\emph{N}$LG (2$\leq$$N$$\leq$6). Fig. 2(c) shows the optical contrast of the t2LG, t3LG and t4LG in this CVD-MLG flakes as an example. Just one adsorption peak is observed at $\sim$640 nm in the t$\emph{N}$LG if we compare them with those of the corresponding AB-$N$LG. This suggests that there only exists one twisted interface in CVD-grown t$N$LG for $N>2$, and the twisted interface should be the same as that of CVD-grown t2LG (i.e., t(1+1)LG) in the same flake. Therefore, the CVD-grown t$\emph{N}$LG can be identified as t(1+$n$)LG ($n$=$N$-1). To further confirm the above conclusion, the Raman spectra of the CVD-grown t\emph{N}LG are measured, as shown in Fig. 2(d). A couple of R and R$'$ peaks are observed in 1509 and 1621 cm$^{-1}$ in the CVD-2LG, which confirms that this CVD-2LG is a t(1+1)LG with 11.5$^{\circ}$. For the CVD-3LG, only the same couple of R and R$'$ peaks are observed, indicating no more twisted interface in this flake, which means that this CVD-3LG is a t(1+2)LG. Furthermore, the CVD-4LG, CVD-5LG and CVD-6LG are t(1+3)LG, t(1+3)LG and t(1+5)LG, respectively. Optical contrast and R/R$'$-related Raman spectra of other CVD-MLG flakes also reveals their stacking way of t(1+$n$)LG under the present growth condition.

\begin{figure*}[!htb]
\centerline{\includegraphics[width=120mm,clip]{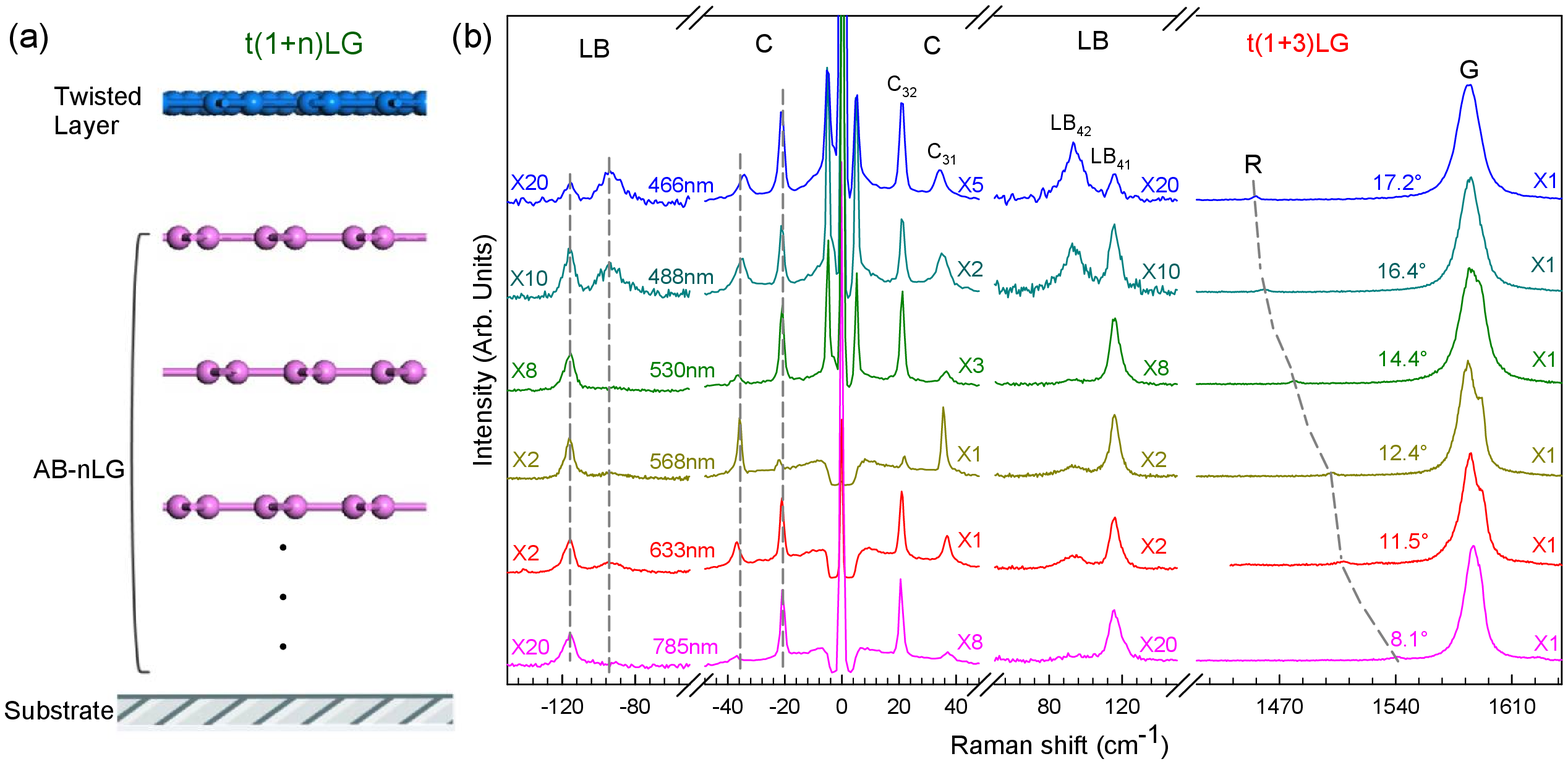}}
\caption{\textbf{Schematic diagram of CVD-grown t(1+$n$)LG and Raman spectra of CVD-grown t(1+3)LG with different twist angles.} ({\bf a}) Schematic diagram of t(1+$n$)LG. ({\bf b}) Stokes/anti-Stokes Raman spectra in the C and LB peak region and Stokes spectra in the R and G peak regions for CVD-grown t(1+3)LGs. $\theta_t$ of each t(1+3)LG are indicated by the peak position of the corresponding R mode. The resonance excitation wavelength ($\lambda$) corresponding to each $\theta_t$ is marked. The spectra are scaled and offset for clarity. Vertical dashed lines are a guide to the eye.} \label{Fig4}
\end{figure*}

The stacking configuration of ME-tMLGs had been investigated by the C and LB modes,\cite{Wu_LBM_ACSnano,wu-nc-2014-resonant} which can only be observed within the resonant excitation window. The 488 nm laser is used to excite the Raman spectra of the CVD-grown t(1+$n$)LG with a $\theta_t$ of 16.9$^{\circ}$, as shown in Fig. 3. For comparison, the Raman spectrum of ME-t(1+1+1)LG excited by 633nm is plotted too. In an AB-$N$LG, where $N$ is the layer number, there are $N$-1 C and LB modes, which are denoted as ${C}_{NN-i}$ and LB$_{NN-i}$ ($i=1,2,...,N-1$), respectively. Here, ${C}_{N1}$ and LB$_{N1}$(i.e., $i=N-1$) are the C and LB modes with the highest frequencies, respectively. All the frequencies of the C and LB modes with different layer numbers can be calculated by a linear chain model (LCM), where each graphene layer is considered as one ball to calculate the frequency of the C and LB modes and only nearest-neighbor interlayer interactions are taken into account.\cite{tan-2012-NM-shear,zhang-PhysRevB-2013,wu-nc-2014-resonant,Wu_LBM_ACSnano} For t(1+2)LG, just one C peak is observed in $\sim$28 cm$^{-1}$, identified as C$_{21}$, because it's close to the frequency of C mode in AB-2LG\cite{tan-2012-NM-shear}, 31 cm$^{-1}$. According to the peak position, the peak in the LBM region is identified as LB$_{31}$. Although the total layer number of ME-t(1+1+1)LG is the same as that of the CVD-grown t(1+2)LG, only the LB$_{31}$ is the observed at the low frequency zone in t(1+1+1)LG. Because both the interfaces in ME-t(1+1+1)LG are the twisted ones and the twisted interface would obstruct the interlayer shear coupling\cite{wu-nc-2014-resonant,Wu_LBM_ACSnano}, the C mode is absent in its Raman spectrum. Indeed, the C modes in tMLGs are localized in the AB-stacked constituent\cite{wu-nc-2014-resonant,Wu_LBM_ACSnano} However, the twisted interface wouldn't affect the interlayer breathing coupling, wherefore the LBMs are contributed from all the graphene layers.\cite{Wu_LBM_ACSnano} Therefore, it is reasonable that only the LB$_{31}$ can be observed in the ULF region of ME-t(1+1+1)LG. Furthermore, the C modes of 3LG, C$_{31}$ and C$_{32}$ are observed in the t(1+3)LG, and the LBMs of 4LG, LB$_{41}$ and LB$_{42}$, can be detected at the same time. Overall, in the t(1+$n$)LG, the C modes of $n$LG and the LBMs of (1+$n$)LG are observed, confirming the identification by optical contrast and the R and R$'$ modes. Therefore, combining with high-frequency R/R$'$ modes and ULF C/LB modes, all the stacking configuration information, including layer number of each constituent, the stacking way between two adjacent constituents and twist angle of each twisted interface, of one tMLG flake can be identified.

During the growth process of graphene layers, the C atoms diffuse from the nuclear of the first layer to the substrate to form the second layer, and so on to form a MLG. The twist stacking is more likely to form between the top two layers, due to the influence of the copper foil. However, the relative orientation between the layers below the top one is more likely to be AB-stacked. The schematic diagram of CVD-grown t(1+$n$)LG is demonstrated in Fig. 4(a). It is noteworthy that the LBMs are more difficult to be observed than the C modes in the AB-MLG\cite{tan-2012-NM-shear,lui2014temperature}, and only the C$_{N1}$ were observed in the Raman measurement at room temperature, due to the symmetry and weak electron-phonon coupling.\cite{tan-2012-NM-shear,lui2014temperature} Here, LBM and more than one C mode are detected in CVD-grown t(1+$n$)LG flake with high signal to noise ratio, because the resonance from the VHS in JDOS enhances the Raman intensity and the twisting between two constituents leads to lower symmetry, which makes more C and LB modes Raman active.\cite{wu-nc-2014-resonant,Wu_LBM_ACSnano} In the graphite and ME-MLG, the AB-stacked MLG is the most stable stacking order and the ABC stacked ones account for only about 15\%.\cite{lui-2010-NL-imaging,lipson1942structure} The C and LB modes of CVD-growth t(1+n)LG are also in analog to those in ME-t(1+n)LG, where nLG is with AB stacking. Furthermore, the transition temperature of ABC-AB modification in bulk graphite is higher than 1000 $^o$C, however, the ABC-AB transition in ME-MLG has been observed within two months even at room temperature, and thus it is expected that the transition temperature of ABC-stacked MLG will be much lower than that in bulk case. Additionally, the growth temperature of the CVD-MLG is as high as 1020 $^o$C. Therefore, it's reasonable to conclude that the constituent nLG in the observed t(1+n)LGs is with AB stacking.

\begin{figure*}[!htb]
\centerline{\includegraphics[width=120mm,clip]{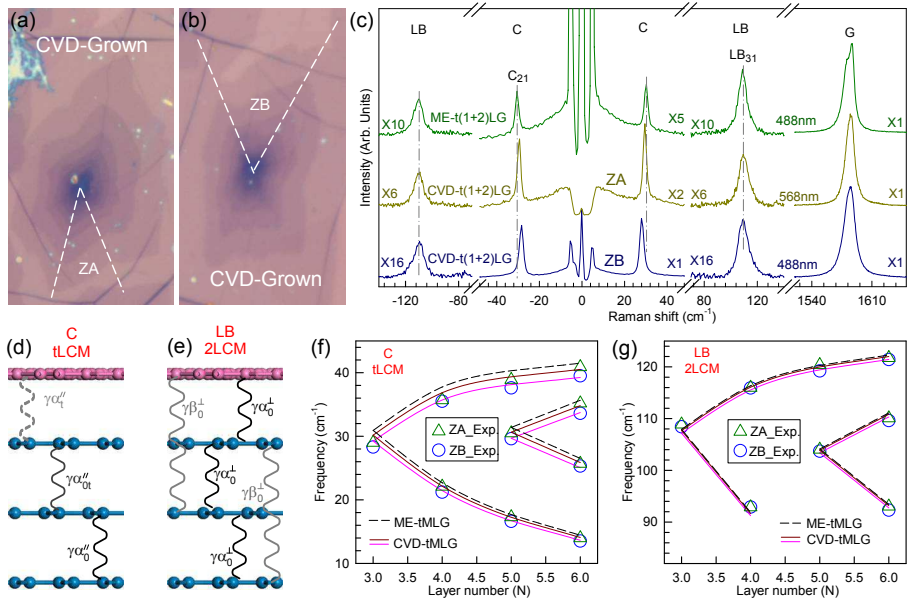}}
\caption{\textbf{Interlayer coupling of ME-t(1+$n$)LG and CVD-grown t(1+$n$)LG.} ({\bf a}) Optical image of one CVD-MLG flake. One of the monocrystalline zones is marked as ZA. ({\bf b}) Optical image of another CVD-MLG flake. One of the monocrystalline zones is marked as ZB. ({\bf c}) Stokes/anti-Stokes Raman spectra in the C and LB peak region and Stokes spectra in the G peak region for ME-t(1+2)LG (green) and the two CVD-grown t(1+2)LGs (ZA and ZB). Vertical dashed lines are a guide to the eye. ({\bf d}) The schematic diagram of tLCM for calculating the C frequencies in tMLGs. ({\bf e}) The schematic diagram of 2LCM for calculating the LBM frequencies in tMLGs. ({\bf f}) Summary of experimental frequencies of the C modes (green triangle for ZA and blue circle for ZB). The calculated frequencies of the C modes are plotted by lines too (dash line for mechanical exfoliation, dark red solid line for ZA and pink solid line for ZB). ({\bf g}) The experimental and calculated LB peak frequencies for each sample.}
\label{Fig5}
\end{figure*}

To comprehensively figure out the stacking configuration of t(1+$n$)LG grown by the present CVD method, more tMLG flakes with different $\theta_t$ are detected with the corresponding resonant excitation energies, as shown in Fig. 4(b). The t(1+3)LG in each tMLG flake is selected for comparison. $\theta_t$ of these six t(1+3)LGs are distributed from 8.1$^{\circ}$ to 17.2$^{\circ}$, which can be determined by the position of the R peak marked by the gray dash line. For each t(1+3)LG, the C$_{31}$ and C$_{32}$ modes are observed at $\sim$36 cm$^{-1}$ and $\sim$21 cm$^{-1}$, respectively, which are close to the corresponding peak position of ME-t(1+3)LG ($\sim$37 cm$^{-1}$ and $\sim$22 cm$^{-1}$, respectively)\cite{wu-nc-2014-resonant}. Moreover, for all t(1+3)LGs, the LB$_{41}$ and LB$_{42}$ are measured in $\sim$116 cm$^{-1}$ and $\sim$93 cm$^{-1}$, which are almost the same as those of ME-t(1+3)LG\cite{Wu_LBM_ACSnano}. These results indicate that the AB-stacked between the layers below the top layer don't depend on the twist angle of the top two layers, and the t(1+$n$)LG is the universal stacking configuration of our CVD-MLGs under the present growth condition. The C and LB modes in all the t(1+3)LGs exhibit similar spectral features of the C and LB modes to the ME-t(1+3)LG\cite{Wu_LBM_ACSnano}. This means that the CVD method can be used to produce tMLG with good interlayer coupling, and tMLG can survive with its stacking configuration during the transferring process from copper foil to the Si/SiO$_{2}$ substrate.

The C and LB modes are the direct signature of the interlayer coupling. The strength of the interlayer coupling can be determined from the position of the C and LB modes. To understand the interlayer coupling quantitatively, the frequency of C and LB modes in CVD-MLG flakes is compared with that of the corresponding ME-MLG. As shown in Figs. 5(a)(b), the optical images of two CVD-MLG flakes are shown. The monocrystalline zone of each flake is marked as ZA and ZB, respectively. The Raman spectra of ME-t(1+2)LG and two CVD-grown t(1+2)LGs (from ZA and ZB) are presented in Fig. 5(c). LB$_{31}$ of those t(1+2)LG are almost located at same position, $\sim$109 cm$^{-1}$. However, C$_{21}$ mode of the ME-t(1+2)LG is at $\sim$30.5 cm$^{-1}$, which is higher than that of the two CVD-t(1+2)LGs, $\sim$29.6 cm$^{-1}$ and 28 cm$^{-1}$ from ZA and ZB, respectively.

For the 2D layered materials, according to the LCM, the frequencies of interlayer vibration modes are proportional to the square root of the interlayer force constant, which is usually assumed as $\alpha$ and $\beta$. In the ME-tMLG system, a softened factor with respect to the bulk case is introduced at the twisted interface for shear coupling, ${\alpha}_{t}^{\parallel}/{\alpha}_{0}^{\parallel}\sim$0.2, and another soften factor with respect to the bulk case is considered at the AB-stacked interface next to the twisted interface, ${\alpha}_{0t}^{\parallel}/{\alpha}_{0}^{\parallel}\sim$0.9. This LCM with a weakened coupling at the twisted interface is known as tLCM.\cite{wu-nc-2014-resonant} For the LBMs, the layer breathing coupling (${\alpha}_{0}^{\perp}$) at the twisted interface is as strong as that at the AB-stacked interface. However, to well predict the experimental frequencies, the second-nearest layer breathing coupling (${\beta}_{0}^{\perp}$) must be considered in the LCM, given 2LCM.\cite{Wu_LBM_ACSnano} tLCM and 2LCM, as shown in Figs. 5(d)(e), predict the frequencies of C and LB modes very well in the ME-tMLG respectively, whose results are plotted in Figs. 5(f)(g) by dash black lines. tLCM and 2LCM are also adopted to describe the interlayer vibration modes of CVD-grown samples. The C and LB modes frequencies of each CVD-tMLG are summarized in Figs. 5(f)(g). Compared with the frequencies in ME-tMLG, there are the softening of the experimental frequency for all the whole-family modes in CVD-tMLG. Therefore, a unified softened factor $\gamma$ for each interlayer coupling in tLCM and 2LCM is assumed, as shown in Figs. 4(d)(e). For the C modes, $\gamma$ is assumed as 95\% and 90\% for ZA and ZB, respectively, to fit the experimental results, which means that the interlayer shear coupling is softened slightly, which demonstrates the high quality of these CVD-grown tMLGs. However,For the LBMs, $\gamma$ is 99.5\% and 99\% for ZA and ZB, respectively, which indicates that the interlayer breathing coupling almost keep constant as the bulk case comparing with the ME-tMLGs, as shown in Fig. 5(g).

In general, there are two main reasons for the softening of intrinsic interlayer coupling in CVD-grown tMLGs: the change of interlayer space changing resulting from molecules intercalation and wrinkles, and the change of stacking way changing resulting from twisting. The larger interlayer space would significantly soften the frequency of both C and LB modes. In the present case, the LBM frequency almost keep constant as the bulk case, independent on the twist angle, which means the interlayer space almost keep constant as the bulk case, independent on the twist angle. Moreover, twisting between two adjacent layers would just affect the shear coupling at the twisted interface. However, the softening happens for all C modes, which means the twisting isn't the reason of lower C modes in CVD-grown tMLGs. The CVD-grown tMLGs are with a bit of defects, like atom missing or dislocation.\cite{li-2009-large,kim_2011_grain_acsnano} These defects would affect the alignment of carbon atoms of neighboring layer at the defect region. The shear coupling are much more sensitive to the alignment of carbon atoms than layer breathing coupling.\cite{Wu_LBM_ACSnano,puretzky2015low} Therefore, the C and LB modes also can be used to evaluate the quality of CVD-grown tMLGs.

\section{Conclusion}

In conclusion, we prepared the MLGs by CVD method. The contrast mapping of Rayleigh scattering is employed to identify the layer number. The CVD-MLG flakes can be distinguished with different
zones by the Raman mapping of the enhanced intensity of the G mode excited by a specific laser energy. Only one R mode is observed in each zone containing different MLGs, which suggests that the MLG flakes are twisted ones as t(1+$n$)LGs. For the CVD-grown MLGs with a definite layer number, they exhibit similar spectral features, including the number and mode frequency of the C and LB modes, independent of the twist angle, confirming the universality of the stacking configuration in the CVD-grown tMLGs under the present growth condition. The frequencies of the C and LB modes in the CVD-grown MLGs are almost identical to those in the corresponding ME-MLGs with the same total layer number and sub-constituents, indicating the high crystal quality of the CVD-grown MLGs.

\section{Acknowledgments}
We acknowledge support from the National Basic Research Program of China Grant Nos. 2016YFA0301200 and 2014CB932500, and the National Natural Science Foundation of China, grants 11225421, 11434010, 11474277, 11504077 and 21525310.

\bibliographystyle{carbon}
\bibliography{twisted_CVD}
\end{document}